\documentclass[11pt]{article}
\usepackage{graphicx}
\usepackage{amssymb,amsmath,array}
\begin{document}
\title{
A Silicon-Tungsten ECal with Integrated Electronics} 
\author{R. Frey$^1$, J. Brau$^1$, M. Breidenbach$^2$, D. Freytag$^2$,
G. Haller$^2$, R. Herbst$^2$, R. Lander$^3$, \\ 
T. Nelson$^2$, V. Radeka$^4$, D. Strom$^1$, and M. Tripathi$^3$
\thanks{This work supported in part by the U.S. DoE LCDRD program.}
\vspace{.3cm}\\
1- University of Oregon - Dept of Physics and Center for High Energy Physics \\
Eugene, Oregon - USA
\vspace{.1cm}\\
2- Stanford Linear Accelerator Center \\
Stanford, California - USA
\vspace{.1cm}\\
3- University of California - Dept of Physics \\
Davis, California - USA
\vspace{.1cm}\\
4- Brookhaven National Laboratory - Physics Division \\
Upton, New York - USA
}

\maketitle
\vspace{-0.2cm}
{\small\em To appear in Proceedings of the Linear Collider Workshop 2007,
DESY, Hamburg, Germany, June 2007.}

\begin{abstract}
We summarize recent R\&D progress for a silicon-tungsten electromagnetic calorimeter
(ECal) with integrated electronics, designed to meet the ILC physics requirements.
\end{abstract}

\section{Overview}

\begin{figure}[htb]
\begin{center}
\includegraphics[width=10cm]{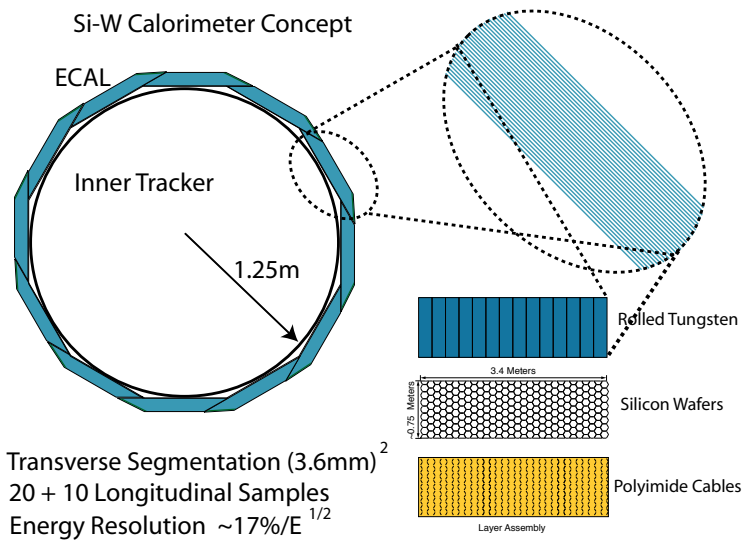}
\end{center}
\caption{Silicon-tungsten ECal as envisioned for the SiD concept.}
\label{Fig:frey_concept}
\end{figure}

A basic physics requirement for ILC detectors is that they provide excellent 
reconstruction of hadronic final states. This allows access to new physics 
which is complementary to the LHC.
One statement for a requirement on jet reconstruction is that intermediate 
particles which decay into jets, such as W, Z, or top, can be identified and isolated. 
This places unprecedented requirements on 2-jet or 3-jet mass resolution, 
typically at the level of 3-5\% using the PFA technique, which makes challenging 
demands on the calorimeters. The electromagnetic energy resolution is not expected 
to limit jet resolution using a PFA. However, particle separation---photon-photon 
and charged hadron-photon---is crucial. In addition, if one provides this kind of imaging calorimeter to meet the PFA needs, these same features will also be put to good 
use for reconstruction of specific tau decay modes (to enable final-state polarization measurement), to ``track'' photons (even if originating from a vertex displaced
from the interaction point), to track MIPS, and so forth. 
Figure \ref{Fig:frey_concept} and Table \ref{Tab:frey_parameters} provide
some context for our ECal design within the SiD detector concept, along with
some main design parameters. More detail is included in the presentation\cite{frey_url}.

\begin{table}[htb]
\caption{Main parameters of the silicon-tungsten ECal for SiD.}
\label{Tab:frey_parameters}
\begin{center}
\vspace{-0.3cm}
\begin{tabular}{|l|c|}
\hline
inner radius of ECal barrel & $1.27$ m \\ \hline
maximum z of barrel & $1.7$ m \\ \hline
longitudinal profile & (20 layers $\times$ $0.64 X_0$ ) $+$ (10 layers $\times$ $1.3 X_0$ )\\ \hline
silicon sensor segmentation & 1024 hexagonal pixels \\ \hline
pixel size & 13 mm$^2$  \\ \hline
readout gap & 1 mm (includes $0.32$ mm silicon thickness)\\ \hline
effective Moliere radius & 13 mm \\ \hline
pixels per readout chip & 1024 \\ \hline
\end{tabular}
\end{center}
\end{table}

The thrust of our R\&D project is to integrate detector pixels on a large, 
commercially feasible silicon wafer, with the complete readout electronics, 
including digitization, contained in a single chip (the {\em KPiX} ASIC) 
which is bump bonded to the 
wafer. We take advantage of the low beam-crossing duty cycle ($10^{-3}$) to reduce 
the heat load using power cycling, thus allowing passive-only thermal management. 
Our design then has several important features:
The electronics channel count is effectively reduced by a factor of 1024;
the transverse segmentation down to a few mm can be naturally accommodated
(with the cost, to first order, not dependent on the segmentation choice);
the readout gaps can be small (1 mm). 
This last property is crucial for maintaining the small 
Moliere radius intrinsic to tungsten.

\section{Sensor and electronics progress}

\begin{figure}[htb]
\begin{center}
\includegraphics[width=8.5cm]{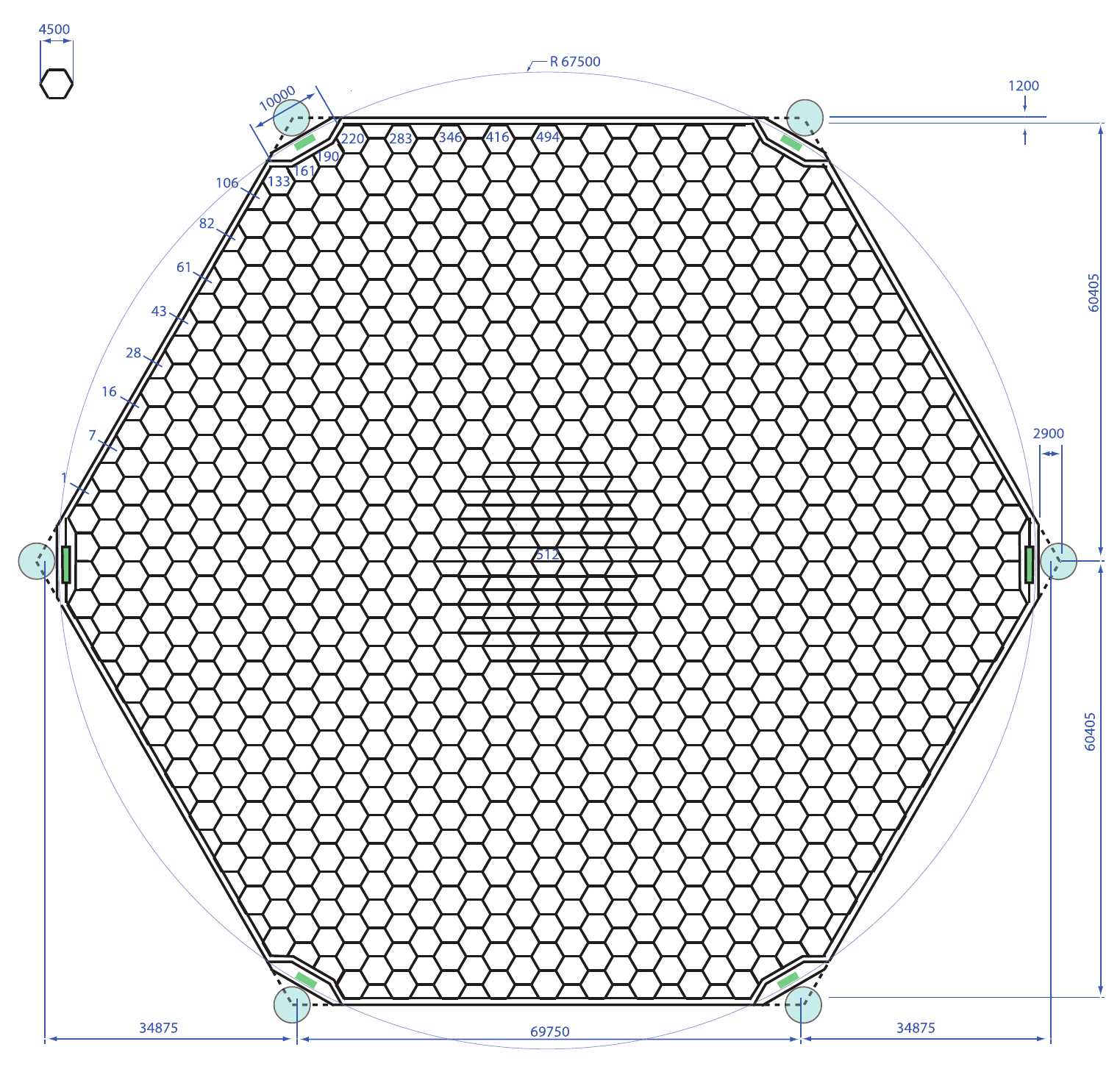}
\caption{Schematic of version 2 silicon sensors. The central region includes
a pad array to which the KPiX ASIC is to be bump or gold-stud bonded.}
\end{center}
\label{Fig:frey_det2}
\end{figure}

Based on the lab measurements\cite{strom_det1} performed on the version 1 silicon 
sensor prototypes, we have developed a design for new (version 2) sensors which
can be used to fabricate a full-depth (30-layer) ECal module.
The new sensor design is depicted in Fig.~\ref{Fig:frey_det2}. The layout minimizes 
capacitive and resistive noise contributions from the signal traces, especially
in the vicinity of the KPiX chip. A typical trace contributes $C\sim 20$ pF 
and $R\sim 300$ $\Omega$.

The readout of the Si pixels must accommodate a very large dynamic range. 
Based on EGS4 simulations, the 
largest signals in a single pixel---arising from 500 GeV Bhabha electrons---correspond to about 2000 MIPs at shower max. At the low end, one requires measuring MIPs well above
the elcetronic noise (SNR$\approx 7$ or better). 
The KPiX design incorporates this large dynamic range in a novel way, 
using on-the-fly range switching. 
Figure 3 shows this 
range-switching function in action in the lab. In the plot, as the injected charge is increased, we see the range switch at about 700 fC. For 320 micron silicon, 1 MIP is equivalent to about 4.1 fC. Thus the upper end of the plot corresponds to about 
2500 MIPs, more than the expected maximum. 

\begin{figure}[htbp]
\begin{center}
\includegraphics[width=10cm]{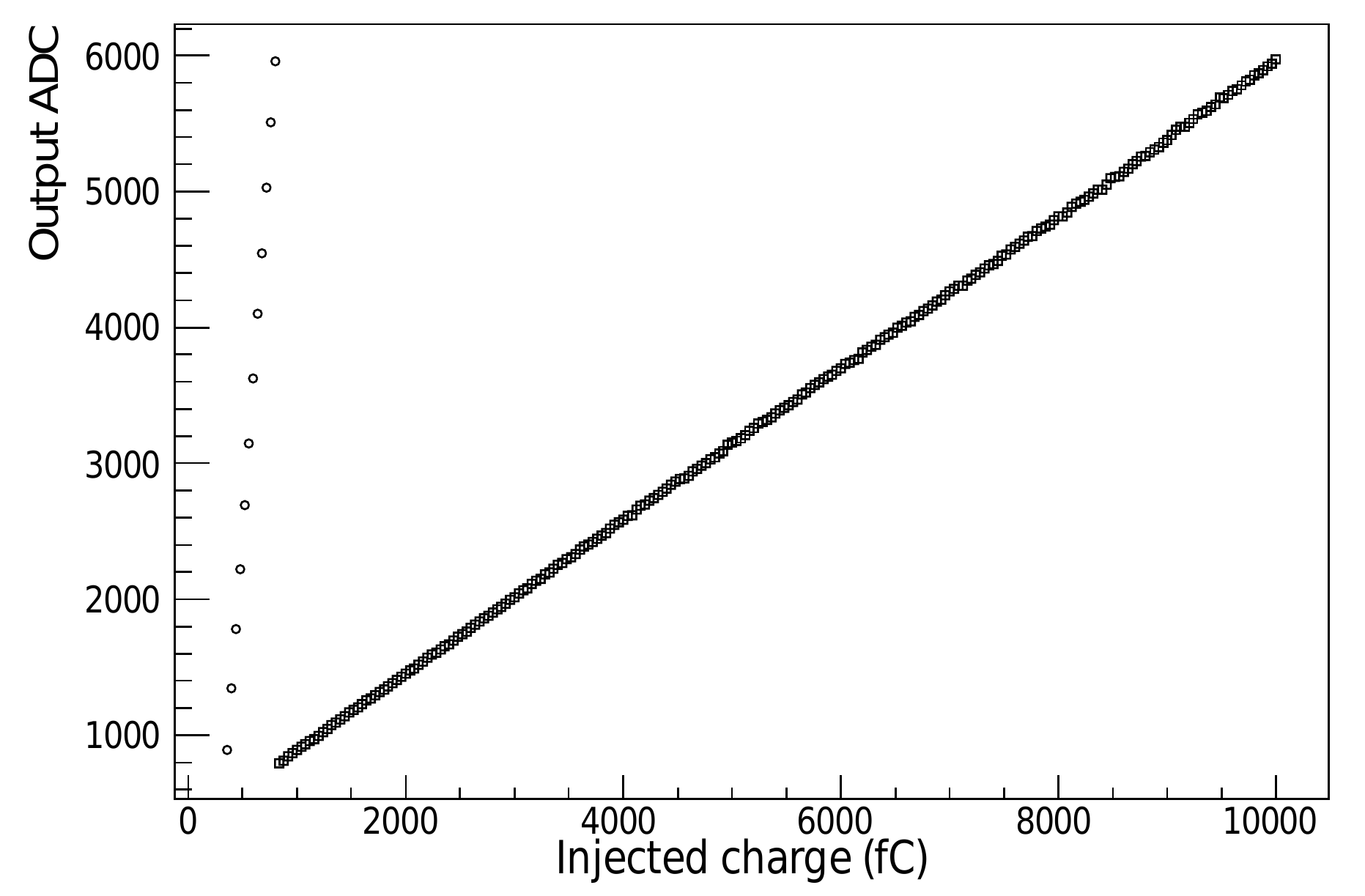}
\caption{KPiX output as a function of input charge, showing the 
dynamic gain change at about 700 fC.}
\end{center}
\label{Fig:frey_range}
\end{figure}

If the KPiX heat load is kept below about 40 mW, the temperature gradient across an 
ECal module ($\approx 75$ cm) can be kept at an acceptable 
level ($<10^\circ$ C) using only passive 
heat conduction via the tungsten radiators. Clearly, this is desirable, and is 
achievable by taking advantage of the beam-timing structure of the ILC, where beams 
are only present for 1 ms out of each 200 ms cycle. By cycling off most of the
(analog) KPiX power between beam-crossing times,
lab measurements of the prototypes confirm that the heat load of the 
full KPiX chip will be about 20 mW.
This heat is to be passively conducted to the module edges behind the ECal (see Fig.~\ref{Fig:frey_concept}), where it can be extracted.

Recently, a KPiX v4 prototype was connected to a spare CDF silicon-strip
detector and placed in a test beam at the SLAC End Station A (ESA).
The data set is still being analyzed. A preliminary result is given in
Fig.~4,
which shows the detected charge distribution.
Since the beam rate averaged $0.25$ electron per pulse, and the beam
diameter was much larger than the active detector region, the distribution
is dominated by the electronic noise, which is gaussian over several orders of
magnitude. The single-MIP contribution is clearly visible. 

\begin{figure}[htbp]
\begin{center}
\includegraphics[width=13cm]{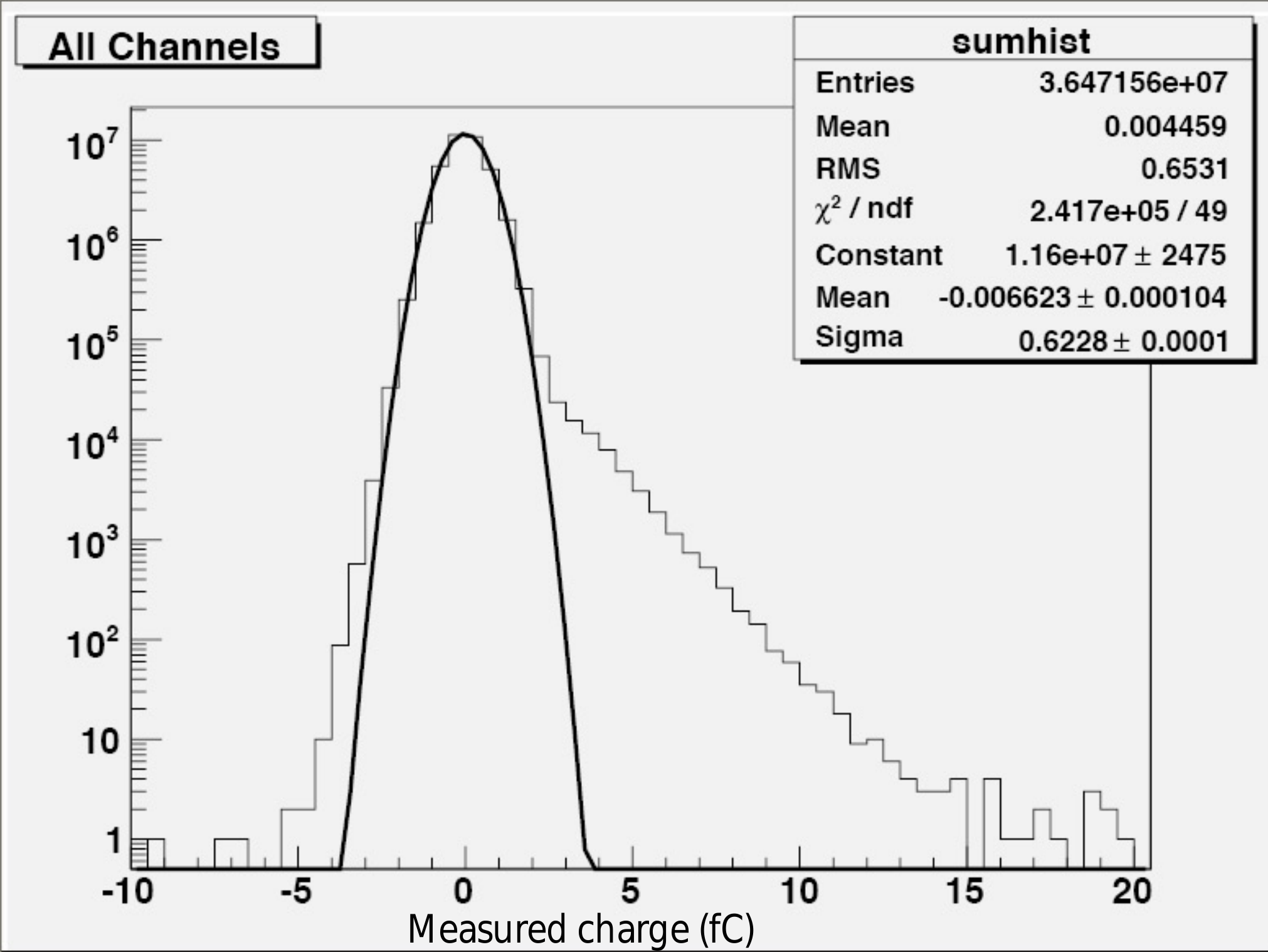}
\caption{Distribution of charge collected by the KPiX v4 chip
at the SLAC ESA.}
\end{center}
\label{Fig:frey_beam}
\end{figure}


\begin{footnotesize}


\end{footnotesize}


\end{document}